\documentclass[10pt]{article}

\usepackage[english]{babel}  

\usepackage{amsmath,amsthm,amsfonts,amssymb}
\usepackage{cmap}
\usepackage[dvips]{graphicx}
\usepackage{wrapfig,epsfig,epsf}
\usepackage{natbib}

\begin{document}
\title{Transient dynamics of large scale vortices in Keplerian disk}
\author{D.N. Razdoburdin$^{1,2}$\thanks{E-mail: d.razdoburdin@gmail.com}}
\date{$^{1}$Department of Physics, Moscow M.V. Lomonosov State University, Moscow, 119992, Russia\\
$^{2}$Sternberg Astronomical Institute, Moscow M.V. Lomonosov State University, Universitetskij pr., 13, Moscow 119992, Russia
}

\maketitle

\begin{abstract}
The mechanism of transition from laminar state to turbulent state in Keplerian disks is still unknown.
The most popular version today is generation of turbulence due to magnetorotational instability (MRI).
However magnetohydrodynamic simulations give the value of Shakura-Sunyaev parameter more then an order of magnitude smaller rather than that found from observations.
One way to solve this problem is the existence of an alternative or additional mechanism for generating turbulence.
It can be the bypass mechanism, which is responsible for transition to turbulence in Couette and Poiseuille flows.
This mechanism is based on the transient growth of linear perturbations in the flow with the subsequent transition to the nonlinear stage.

In order to clarify the role of this mechanism in astrophysical disks first of all it is necessary to calculate the maximal possible growth factor of linear perturbations in the flow.
In this paper the results of such calculations are presented for perturbations on different scales compared with the disk thickness.
Qualitative description of mechanisms responsible for the growth will also be presented.
It was found that the most rapidly growing shear harmonics have azimuthal wavelength of the order of the disk thickness.
In addition, their initial form is always similar to the vortex perturbations with the same potential vorticity.
It was shown that the vortices with azimuthal wavelength more than an order of magnitude in excess of the disc thickness, are still able to grow by dozens of times.
\end{abstract}

\begin{keywords}
hydrodynamics --- accretion, accretion discs --- instabilities --- turbulence
\end{keywords}

\begin{section}{Introduction}
The mechanism which is responsible for energy transfer from the main flow to perturbations must be linear.
This is clear from general conclusion that while perturbations are small all orders of magnitude  are suppressed with respect to the linear as well as from the rigorous derivation for vortical motion (see \cite{schmid-henningson-2001}, \cite{henningson-reddy-1994}).

For this reason, the study of turbulence generation in astrophysical disks started from finding linear perturbations which are growing exponentially with time.
Today we know of two linear instabilities generating turbulence in the disks: magnetorotational instability and vertical shear instability.
The first one owes its existence to the magnetic field, and the second one - to the vertical gradient of the angular rotation velocity of flow.

The role of magnetorotational instability was first shown in the article by \cite{balbus-hawley-1991}.
It has been demonstrated that instability with the maximum increment of $\sim 0.75 \Omega^{-1}$ exists in a magnetized differentially rotating flow, and it generates turbulence in the flow on the nonlinear stage.

Numerical simulations, which have been carried out, for example, in \cite{simon-2012}, predict a value of Shakura-Sunyaev parameter (\cite{shakura-sunyaev-1973}) $\alpha \sim 0.03$.
However, there are some observational evidences (\cite{suleimanov-2008}, \cite{king-2007}, \cite {kotko-lasota-2012}, \cite {lipunova-2015}), that indicate that the value of this parameter in real disks is higher by more than an order of magnitude compared to that obtained in simulations.
Thus the possibility of explanation of the observed effective viscosity by MRI-produced turbulence is still not clear.

Moreover, MRI cannot work in the so-called "dead zones" of protoplanetary disks in which medium ionisation degree is too low for the required coupling with the magnetic field.

The second supercritical instability, which provides the transition to turbulence on the nonlinear stage is the so-called vertical shear instability which is similar to the Goldreich-Schubert-Fricke (GSF) instability.
It was discovered in the work \cite{goldreich-schubert-1967} and independently in the work \cite{fricke-1968}, adapted to the accretion disks in the works \cite{urpin-brandenburg-1998}, \cite{urpin-2003} and \cite{arlt-urpin-2004}.
The possibility of turbulence generation with effective viscosity $\alpha \sim 10^{-3}$ by this instability was shown in the work of \cite{nelson-2013}.
Because this instability is not associated with magnetic field, it can work in "dead zones", but effective viscosity produced by this instability is much less then produced by MRI.

A similar problem existed in laboratory hydrodynamics.
Taylor-Couette flow is linearly stable up to Reynolds numbers less then $Re\sim 5772$, but becomes turbulent already at $Re<350$.
Solution of this problem was connected to the mathematical properties of dynamical operator, which controls linear perturbations in the flow. Because of the significant radial gradient of the flow angular velocity, operator has eigenfunctions which are nonorthogonal.
For this reason, even in linear steady flows, perturbations showing transient growth of amplitude can exist.
This growth provides the energy transfer from regular to turbulent motions in Taylor-Couette flow (see, for example, \cite{trefethen-1993}).
The mechanism associated with transition to turbulence due to transient growth is called bypass mechanism.

And the question about the possibility of the application of this mechanism to accretion and protoplanetary disks of course arises.
The first step is to investigate the opportunity of significant transient growth of linear perturbations in astrophysical disks.
\end{section}

\begin{section}{Transient growth of perturbations in Keplerian disk}
To date the investigation of transient growth of perturbations in astrophysical systems has been described in many papers.
In works \cite{chagelishvili-1997}, \cite{chagelishvili-2003}, \cite{tevzadze-2003}, \cite{afshordi-2005}, \cite{bodo-2005}
\cite{tevzadze-2008}, \cite{heinemann-papaloizou-2009a}, \cite{heinemann-papaloizou-2009b}, \cite{tevzadze-2010}, \cite{volponi-2010}, \cite{salhi-pieri-2014} transient growth of shearing harmonics in spatially local case was investigated.
Transient effects without changing to shearing harmonics but, instead, looking for a linear combination of eigenfunctions that shows the highest possible growth of energy was studied in articles \cite{yecko-2004}, \cite{mukhopadhyay-2005}, \cite{zhuravlev-shakura-2009}, \cite{razdoburdin-zhuravlev-2012}.

In the present paper a fairly new method of investigation of transient dynamics is described.
It was first used in astrophysical study in \cite{zhuravlev-razdoburdin-2014}, but previously had been extensively used in hydrodynamics in papers \cite{luchini-2000}, \cite{corbett-bottaro-2001}, \cite{guegan-2006}, see also \cite{schmid-2007} and \cite{gunzburger-2003}.

To describe dynamics of small perturbations in the disk which is rotating with angular velocity  $\Omega(r)$ with surface density profile $\Sigma(r)$, the sound speed profile $a_{eq}(r)$ in the equatorial plane and polytropic equation of state with the polytropic index $n$ the following system of linearized Euler equations and the continuity equation  were used (see. \cite{zhuravlev-razdoburdin-2014}):
\begin{equation}
\label{sys_A_1}
\frac{\partial \delta v_r}{\partial t} = -{\rm i}m\Omega\, \delta v_r + 2\Omega \delta v_\varphi - \frac{\partial \delta h}{\partial r},
\end{equation}

\begin{equation}
\label{sys_A_2}
\frac{\partial \delta v_\varphi}{\partial t} =  -\frac{\kappa^2}{2\Omega} \delta v_r  -{\rm i}m\Omega\, \delta v_\varphi -\frac{{\rm i} m}{r} \delta h, 
\end{equation}

\begin{equation}
\label{sys_A_3}
\frac{\partial \delta h}{\partial t} = -\frac{a_*^2}{r\Sigma} \frac{\partial}{\partial r} (r\Sigma \delta v_r) -\frac{{\rm i} m a_*^2}{r} \delta v_\varphi  -{\rm i}m\Omega \,\delta h.
\end{equation}
Here $\delta v_r$, $\delta v_{\varphi}$, $\delta h$ $\propto exp(im \varphi)$ are azimuthal complex Fourier harmonics of Euler perturbation of the radial and azimuthal velocity and enthalpy, $a_*^2 \equiv n a_{eq}^ 2 /(n + 1/2)$, $\kappa^2 = (2 \Omega / r) d/dr (\Omega r ^2) $ - square of epicyclic frequency. Coordinates are expressed in terms of the inner radius of the disk $r_0$, time is expressed in units of inversed Kepler frequency on the inner edge $\Omega ^{-1}(r_0)$.

The system (\ref{sys_A_1}-\ref{sys_A_3}) can be written in operator form:
\begin{equation}
\label{dynamic_equation}
\frac{\partial \mathbf{q}}{\partial t}=\mathbf{Aq},
\end{equation}
where vector $\mathbf {q}$ is equal to $\mathbf {q}=\{\delta v_r,\delta v_{\varphi},\delta h\}$.
Solution of equation (\ref{dynamic_equation}) has the form:
\begin{equation}
\label{q_t}
\mathbf{q}(t)=\mathbf{q}(0)e^{\mathbf{A}t}
\end{equation}

It is necessary to find an initial condition $\mathbf{q}(0)$ for which the functional:
\begin{equation}
\label{functional}
\mathcal{G}\left(t\right)=\frac{||\mathbf{q}(t)||^2}{||\mathbf{q}(0)||^2}
\end{equation}
will reach its maximum value provided that the equality (\ref{dynamic_equation}) is fulfilled.
In order to find such initial conditions the method of Lagrange multipliers is used.
In this case, Lagrangian consists of two components: the goal and the penalty terms which is non-zero only if $\mathbf{q}$ does not obey equation (\ref{dynamic_equation}) (see \cite{corbett-bottaro-2001}, \cite{guegan-2006}, \cite{schmid-2007}):
\begin{equation}
\label{lagrangian}
\mathcal{L}\left(\mathbf{q},\tilde{\mathbf{q}}\right)=\mathcal{G}\left(\mathbf{q}\right)-\int\limits_0^{t}\left(\tilde{\mathbf{q}},\dot{\mathbf{q}}-\mathbf{A(q)q}\right)d\tau.
\end{equation}

Extremum of functional (\ref{functional}) is achieved if variations of the Lagrangian with respect to $\mathbf{q} $ and to $\mathbf{\tilde {q}}$ are equal to zero.
The variation with respect to $\mathbf{\tilde{q}}$ leads to the equation (\ref {dynamic_equation}), and variation with respect to $\mathbf{q}$ leads to equation for $\mathbf{\tilde {q}} $, as well as to links between $\mathbf{\tilde{q}} $ and $\mathbf {q}$ in the initial and final time points (see \cite{zhuravlev-razdoburdin-2014}, \cite{razdoburdin-zhuravlev-2015} for details):
\begin{equation}
\label{conjugated_equation}
\frac{\partial \mathbf{\tilde q}}{\partial t}=-\mathbf{A^{\dag}\tilde q},
\end{equation}
where $\mathbf{A}^{\dag} $ - is the operator conjugated to $\mathbf{A}$.
\begin{equation}
\label{coupling_condition_tau}
\mathbf{\tilde q}(t)=\frac{2}{||\mathbf{q}(0)||^2}\mathbf{q}(t)
\end{equation}
\begin{equation}
\label{coupling_condition_zero}
\mathbf{q}(0)=\frac{||\mathbf{q}(0)||^4}{2||\mathbf{q}(t)||^2}\mathbf{\tilde q}(0)
\end{equation}

Solution of equation (\ref{conjugated_equation}), can also be written through the matrix exponential:
\begin{equation}
\label{tilde_q_t}
\mathbf{\tilde q}(t)=\mathrm{e}^{\mathbf{-A^{\dag}t}}\mathbf{\tilde q}(0)
\end{equation}

\begin{figure}[h!]
\includegraphics[width=1\linewidth]{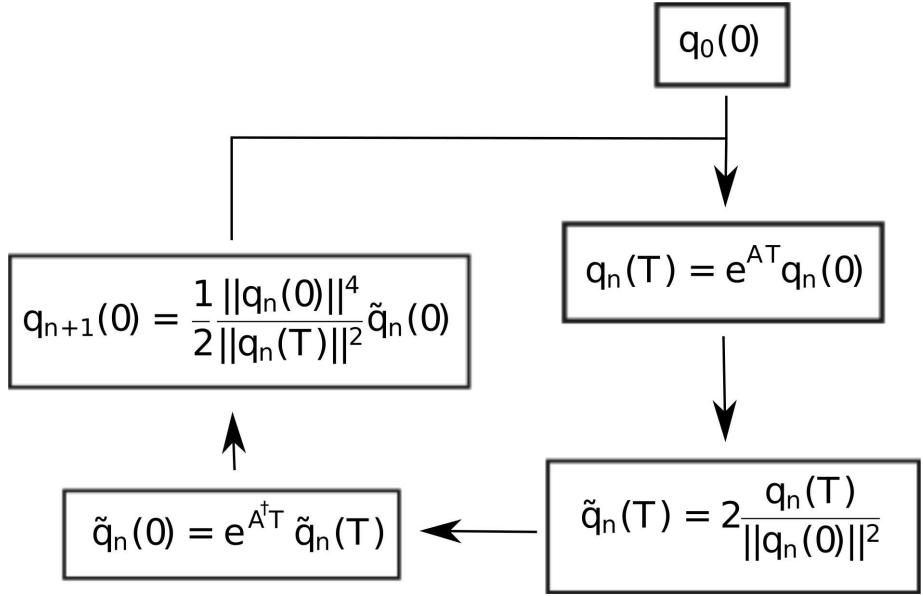}
\caption{\footnotesize{
The scheme of the iteration cycle for finding the optimal perturbations which satisfies the system of the dynamic equations (\ref{dynamic_equation}), see. \cite{schmid-2007}.
}}
\label{scheme}
\end{figure}

To find the explicit form of $\mathbf{A}^{\dag}$ it is nessessary to set a scalar production of vectors.
In this work the scalar production is chosen in such a way that the norm squared of the state vector is equal to the acoustic energy of perturbation:
\begin{equation}
\label{ac_en}
{||{\bf q}(t)||^2}=E_{ac} = \pi \int \Sigma \left ( |\delta v_r|^2 + |\delta v_\varphi|^2  + \frac{|\delta h|^2}{a_*^2}  \right ) r\, dr,
\end{equation}

In this case operator $\mathbf{A}^{\dag}$ corresponds to the following system of equations (\cite{zhuravlev-razdoburdin-2014}):
\begin{equation}
\label{adj_sys_A_1}
\frac{\partial \delta \tilde{v}_r}{\partial t} = -{\rm i}m\Omega\, \delta \tilde{v}_r + \frac{\kappa^2}{2\Omega} \delta \tilde{v}_\varphi - \frac{\partial \delta \tilde{h}}{\partial r},
\end{equation}

\begin{equation}
\label{adj_sys_A_2}
\frac{\partial \delta \tilde{v}_\varphi}{\partial t} =  -2\Omega \delta \tilde{v}_r  -{\rm i}m\Omega\, \delta \tilde{v}_\varphi -\frac{{\rm i} m}{r} \delta \tilde{h}, 
\end{equation}

\begin{equation}
\label{adj_sys_A_3}
\frac{\partial \delta \tilde{h}}{\partial t} = -\frac{a_*^2}{r\Sigma} \frac{\partial}{\partial r} (r\Sigma \delta \tilde{v}_r) -\frac{{\rm i} m a_*^2}{r} \delta \tilde{v}_\varphi  -{\rm i}m\Omega \,\delta \tilde{h},
\end{equation}

The joint solution of the systems (\ref{sys_A_1} - \ref{sys_A_3}) and (\ref{adj_sys_A_1} - \ref{adj_sys_A_3}) allows us to find profiles of the initial perturbation of enthalpy and velocity, which demonstrate the highest possible growth of acoustic energy at time $t$.
To find the joint solution the method of power iterations was used. 
This method consists of the sequential integration of system (\ref{sys_A_1} - \ref {sys_A_3}) forward in time, and the system (\ref{adj_sys_A_1} - \ref{adj_sys_A_3}) backward in time (see. figure \ref{scheme}) up until the desired accuracy of optimal perturbation is achieved.

\begin{figure}[h!]
\includegraphics[width=1.0\linewidth]{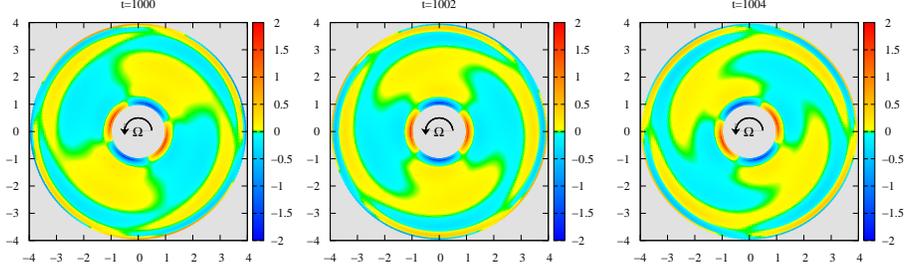}
\caption{\footnotesize{
The contours of the most unstable mode of perturbations with azimuthal wave number equal to $m=2$ in a model of quasi-Keplerian thin disk of finite radial extent.
This model is described in \cite{razdoburdin-zhuravlev-2012}.
The calculation parameters are equal to: characteristic half-thickness of the disk $\delta=0.3$, positions of external and inner boundaries $r_1=1$, $r_2=4$, polytropic index $n=3/2$.
The mode has increment of $\Im[\omega] \approx 0.001$ and phase velocity $\Re[\omega] \approx 0.26$. 
The time is expressed in units of the inversed Keplerian frequency on the inner boundary of the disc. 
It indicates the time elapsed from the starting point at which the mode has unit amplitude.
The arrow shows the direction of substance rotation in the disc.
}}
\label{moda}
\end{figure}

Since this method allows one to find the perturbation which shows the highest possible growth of the energy, it can be used not only for calculation of transient growth, but also for finding the most unstable eigenfunction in a spectraly unstable flow.
An example of such unstable mode is displayed in the figure \ref{moda}.
It is easy to notice that the the mode's pattern rotates rigidly even in a differentially rotating flow.

\begin{figure}[h!]
\includegraphics[width=1.0\linewidth]{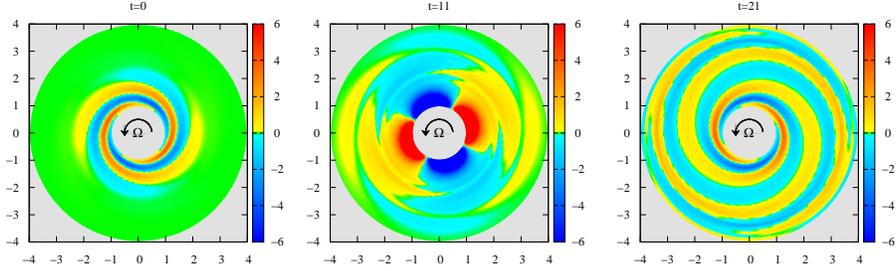}
\caption{\footnotesize{
The contours of the perturbation with $m=2$ that shows the maximum possible transient growth of total acoustic energy at time $t_{opt}=10$.
Times are given in units of the inversed Keplerian frequency on the inner boundary of the disc.
Initial perturbation has conditionally unit amplitude.
Flow model is taken the same as in fig. {\ref {moda}}.
}}
\label{TG}
\end{figure}

In contrast to the mode, the transiently growing perturbation initially represents a leading spiral unwound by the differential rotation of the flow. 
Amplitude of the spiral increases during this process.
After reaching the maximum amplitude, the spiral begins to twist in the other direction, and the amplitude is reduced (see figure \ref {TG}).
In the work by \cite{zhuravlev-razdoburdin-2014} was shown that at the initial time the  form of the optimal perturbation is close to the vortex with the same potential vorticity.

\begin{figure}[h!]
\includegraphics[width=1\linewidth]{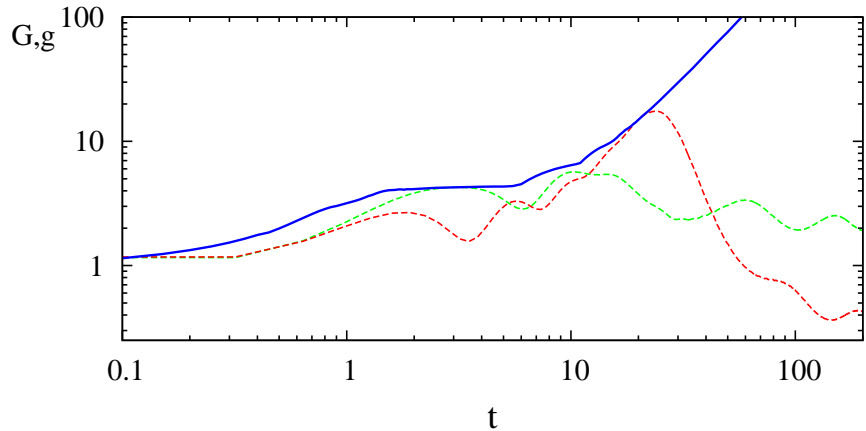}
\caption{\footnotesize{
The optimal growth vs. time.
The solid curve is an optimal growth as the function of time for a disk of Shakura-Sunyaev with $m=5$, $\delta=0.05$, $n=3/2 $.
Dashed curves show the energy of an individual perturbations that are optimal for the instants $t=20$ and $t=3$.}}
\label{G_t}
\end{figure}

To calculate the value of transient growth $G\left(t\right)=\frac{||\mathbf{q}(t)||^2} {||\mathbf{q}(0)||^2}$ the standard Shakura-Sunyaev disk model in the limit of low viscosity was used as a background (\cite{shakura-sunyaev-1973}).
Profiles of the surface density and sound speed can be found in \cite{zhuravlev-razdoburdin-2014}).
In the figure \ref {G_t} the blue curve shows the dependence of the optimal growth with respect to time.
It is easy to notice that the curve is visually divided into three parts: up to $t\sim 2$ $G (t)$ shows quite rapid growth, then up to $t\sim 7$ growth becomes much more slow, and when after $t\sim 7$ it restores.
This behavior is connected to the fact that at small times perturbations in the form of epicyclic motions are dominated, but long term twisted spirals like the one shown in figure \ref{TG} become dominant.

\begin{figure}[h!]
\includegraphics[width=1\linewidth]{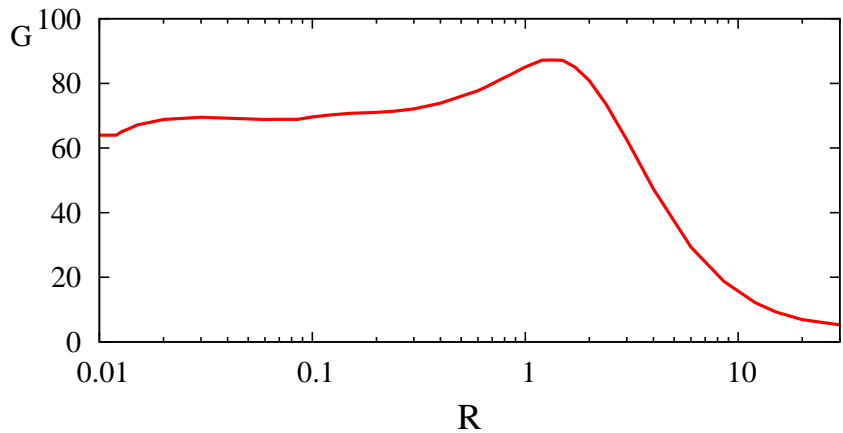}
\caption{\footnotesize{
The dependence of the optimal growth at time $t_{opt}=10$ with respect to azimuthal wavelength for a background flow in form of homogeneous disk described in \cite {zhuravlev-razdoburdin-2014}.
}}
\label{G_R}
\end{figure}

The dependence of the optimal growth on the azimuthal wavelength is shown in figure \ref{G_R}.
For a more convenient parameterization of the problem, the parameter $R$ is used.
It is equal to the ratio of the azimuthal wavelength to the thickness of the disk on the left boundary:
\begin{equation}
R=\frac{\lambda_{\varphi}}{h}=\frac{3/2}{ma_{*}|_{r=1}}
\end{equation}
Perturbations with wavelengths of the order of the disk thickness $R\sim 1$ are amplified most effectively, becouse of the prominent exitation of shearing density waves (\cite{heinemann-papaloizou-2009a}, \cite{zhuravlev-razdoburdin-2014}).
However large-scale vortices with azimuthal wavelength more than an order of magnitude in excess of the disc thickness, are still able to grow by dozens of times.
\end{section}

\begin{section}{Conclusions}
The ability of Keplerian flow linear perturbations to significant transient growth is demonstrated above.
Such growth is a necessary but not sufficient condition for launching the bypass scenario of transition to turbulence.
For transition from growth of linear perturbations to a self-sustaining non-linear turbulence it is necessary to have mechanisms that would translate the trailing spiral similar to that shown in figure \ref {TG} to leading.
Today we can provide arguments both for and against the existence of such mechanism.

On the one hand the effects of tidal and Coriolis forces, that are absent in the Couette flow, dramatically stabilize the flow (see figure 9 in \cite{balbus-hawley-1998}).
This argument is supported by the spatially local numerical simulations \cite{hawley-1999}, \cite{shen-2006} and by the series of laboratory experiments \cite {ji-2006}, \cite {schartman-2009}, \cite {schartman-2012}.
Stability of the quasi-Keplerian flow was observed up to Reynolds numbers $ {\rm Re}= 2\times 10^6$ in this experiments.

On the other hand, due to the extremely low molecular viscosity in astrophysical disks Reynolds number can reach the value of $Re\sim 10^{10}$.
In addition subcritical turbulence observed in flows of cyclonic type (about cyclonic and anticyclonic flows see, for example, \cite {lesur-longaretti-2005}) at the high but finite Reynolds numbers (see \cite {taylor-1936}, \cite {wendt-1933}).
Moreover, the negative results obtained in numerical simulations, can be explained by insufficient numerical resolution (see \cite {longaretti-2002}).
In subsequent work \cite{lesur-longaretti-2005} dynamic of perturbations in the cyclonic and anticyclonic flow was compared by numerical methods.
It was concluded that the requirements for the numerical resolution for the anticyclonic flow is much higher than for the cyclonic, and current numerical resources is insufficient to detect turbulence in Keplerian flow.
Also, the possibility of turbulence generation cannot be eliminated by only the stabilizing action of the Coriolis force.
Finally, another laboratory experiment, which is presented in articles \cite{paoletti-lathrop-2011} and \cite{paoletti-2012}, shows the generation of subcritical turbulence and angular momentum transfer to the periphery of the quasi-Keplerian flow.
The contradictions in the results of the two different groups of researchers, show the complexity of the experiments.

Thus the ability of the bypass scenario in astrophysical disks is the subject of future research.
\end{section}

\section*{Acknowledgements}
The author thanks V.V. Zhuravlev for careful reading the manuscript and important remarks. 
The work is supported by the Russian Science Foundation grant 14- 12-00146.
The equipment used was funded by the Lomonosov Moscow State University Program of Development.

\end{document}